\documentclass[12pt,a4paper,final]{iopart}

\usepackage{iopams}  
\usepackage{graphicx}
\usepackage[breaklinks=true,colorlinks=true,linkcolor=blue,urlcolor=blue,citecolor=blue]{hyperref}
\usepackage{graphicx}
\usepackage{dcolumn}
\usepackage{bm}
\usepackage[]{units}	
\usepackage{textcomp}
\usepackage{eqnarray}

\begin{document}

\title[]{Rotational state detection of electrically trapped polyatomic molecules}

\author{Rosa Gl\"ockner, Alexander Prehn, Gerhard Rempe, and Martin Zeppenfeld} 
\address{Max-Planck-Institut f\"ur Quantenoptik, Hans-Kopfermann-Str. 1, D-85748 Garching, Germany}
\ead{Martin.Zeppenfeld@mpq.mpg.de}

\date{\today}

\begin{abstract}

Detecting the internal state of polar molecules is a substantial challenge
when standard techniques such as resonance-enhanced multi photon
ionization (REMPI) or laser-induced fluorescense (LIF) do not work. As
this is the case for most polyatomic molecule species, we here investigate an
alternative based on state selective removal of molecules from an
electrically trapped ensemble. Specifically, we deplete molecules by
driving rotational and/or vibrational transitions to untrapped states.
Fully resolving the rotational state with this method can be a
considerable challenge as the frequency differences between various
transitions is easily substantially less than the Stark broadening in an
electric trap. However, making use of a unique trap design providing
homogeneous fields in a large fraction of the trap volume, we successfully
discriminate all rotational quantum numbers, including the rotational
$M$-substate.

\end{abstract}

\vspace{2pc}

\maketitle

\tableofcontents

\section{Introduction}

Cold and ultracold molecules offer a large variety of applications in quantum information~\cite{Andre2006,Rabl2006,Wei2011} and quantum simulation~\cite{Pupillo2009,Wall2013a} as well as for high precision measurements~\cite{DeMille08,Hudson2011,Baron2014} or for quantum chemistry and cold collision studies~\cite{Bell2009,Lemeshko2013a}. Triggered by these prospects, an immense effort has been focused on the development of methods for the production of cold and ultracold molecules during the last years. The coldest molecular ensembles have been achieved based on association of ultracold atoms using Feshbach resonances or photoassociation~\cite{Kerman2004,Deiglmayr2008,Ni2008}. However, these techniques involve substantial experimental effort and the molecular species accessible are restricted to dimers composed of laser cooled atoms, thus mainly to alkali dimers. A more general approach is the direct cooling of molecules. Here buffergas cooling~\cite{Weinstein1998,VanBuuren2009,Hutzler2011,Bulleid2013}, deceleration after, e.g., supersonic expansion~\cite{Bethlem1999,Osterwalder2010,Narevicius2008a,Wiederkehr2012,Fulton2004,Merz2012,Gupta1999,Strebel2010,Chervenkov2014}, and velocity filtering~\cite{Rangwala2003,MotschBoosting} are widely applicable. However, these methods are mainly suited to prepare ensembles above temperatures of about \unit[10]{mK}. Prospects for direct cooling to ultracold temperatures have recently appeared in the form of laser cooling of molecules~\cite{Rosa2004,Shuman2010,Hummon2013,Zhelyazkova2014,Barry2014}.

In our group, we have developed an alternative method for direct motional cooling of molecules using optoelectrical cooling~\cite{Zeppenfeld2009,Zeppenfeld2012a}. This Sisyphus-type cooling scheme is in particular suited for polyatomic molecules, meaning molecules that are composed of more than two atoms. It is expected to reach temperatures below \unit[1]{mK} and thus bridge the gap to a regime where evaporative or sympathetic cooling should be possible. To further develop our method, e.g., to add internal state cooling and control, we have to be able to state selectively detect internal states. Unfortunately, commonly used techniques for rotational state detection of molecules such as resonance-enhanced multi photon ionization (REMPI)~\cite{Antonov1978,Twyman2014,Bertsche2010} and laser-induced fluorescence (LIF)~\cite{Kinsey1977} rely on the excitation of electronic states. Especially for polyatomic molecules, the excitation of electronic states, however, can lead to rapid predissociation~\cite{Herzberg1966}, causing an enormous line broadening and thus a loss of state selectivity. In addition, almost all electronic transitions lie in the UV, some in the deep UV and the generation of laser light at these frequencies can be experimentally challenging.  

In this paper we present a detailed investigation of a rotational-state detection technique that is suitable for a large variety of molecular species, especially polyatomic ones. Our method is based on state-selective depletion of trapped molecules. In contrast to previous experiments~\cite{MotschDepletion}, our depletion method does not incorporate electronic excitations but uses instead vibrational and rotational transitions to transfer the molecules from the rotational state of interest to an untrapped state. Moreover, the detection proceeds in a homogeneous-field dc electric trap that allows to spectrally resolve the transitions~\cite{Englert2011}. The long trapping times of more than \unit[10]{s} inside our trap enable us to implement slow depletion processes such as the use of optical pumping via a vibrational mode. A big advantage of our method is that the detection of the molecules themselves can be accomplished by state insensitive techniques such as a quadrupole mass spectrometer. 

Our paper is organized as follows: In section~\ref{sec:Schemes} we first discuss general schemes for detecting the population of rotational states of symmetric top molecules. We mainly focus on two complimentary schemes based either on the use of microwave (MW) transitions alone, or based on a combination of microwave and infrared (IR) transitions. The experimental setup and the implementation of these schemes for the symmetric top molecule $\rm CH_3F$ is described in section~\ref{sec:Exp}.

The key challenge for the experimental realizations of the detection schemes is to resolve all three symmetric top rotational quantum numbers $J,K$ and $M$. Due to a small dependence of the rotational transition frequencies on $K$ and $M$ this cannot be achieved without effort. In particular, to discriminate $K$ and $M$, we need to drive transitions affecting desired states with specific $K$ and $M$ without driving unwanted transitions affecting unwanted states. We therefore carefully investigate the spectral resolution inside our homogeneous-field electric trap in section~\ref{sec:Limit}. Subsequently we consider the position of wanted and unwanted transitions in section~\ref{sec:Sel}. Based on these considerations, in section~\ref{sec:Rates} we examine the dynamics of the depletion using rate equations. Experimental results are shown in section~\ref{sec:Results}. We obtain excellent agreement between the measurements and results from the rate model. Both confirm that while discriminating the $J$ quantum number is relatively easy, discriminating $K$ and $M$ is more difficult. However, using a combination of depletion schemes, in section~\ref{sec:Results.IRD} we demonstrate a state detection which only depends on the population of molecules in states with a single $K$. Finally, in section~\ref{sec:Results.M} we present results for the detection of molecules exclusively populating a single $M$-substate, characterized by single $J,K$ and $M$ quantum numbers.

\section{ Schemes for rotational state detection}
\label{sec:Schemes}

The general idea of all detection methods shown in this paper is to selectively remove molecules in states which we want to detect from the trapped ensemble. The difference in signal of measurements with and without depletion then yields the state selective signal.
We present two methods involving the driving of rotational and/or vibrational transitions to transfer the population to untrapped states. As the molecular parameters (e.g. the rotational constants, or vibrational transition frequencies, or spontaneous decay rates) can vary over a large range, the individual advantages of the two methods can be used for different rotational states or molecular species. 

\subsection{The symmetric top molecule}
\label{sec:Scheme.SymTop}

Before discussing our rotational state detection, we briefly review the properties of symmetric top molecules. The rotational state of a symmetric top molecule is described by three quantum numbers: the total angular momentum $J$, its projection onto the molecular symmetry axis $K$, and the projection on the electric field axis $M$. In the following we will denote the rotational state by $|J;  \mp K ; \pm M\rangle$ with $\mp K$ chosen positive. In the case of no interaction with external fields the rotational energy is given by 
\begin{eqnarray} \label{eq:rot}
E_{J,K}/h=&B_0 J\left(J+1\right)+\left(A_0-B_0\right)K^2   \\
				 & -D_JJ^2(J+1)^2-D_{JK}J(J+1)K^2-D_KK^4 + ... \nonumber
\end{eqnarray}
Here, the first two terms correspond to the rigid rotor approximation with rotational constants $B_0$ and $A_0$ and the last three are small first-order corrections due to centrifugal distortions.

In the presence of an electric field the degeneracy of the $M$ states is lifted according to the first order Stark shift,
\begin{equation}
\Delta \nu = \frac{\mu \mathcal{E}}{h} \frac{KM}{J(J+1)}.
\label{eq:StarkShift}
\end{equation}
Here $\mu$ is the permanent dipole moment and $\mathcal{E}$ is the electric field strength. For typical electric fields in our experiment, this splitting is on the order of tens to hundreds of \unit[]{MHz} and  depends on all three rotational quantum numbers $J$,$K$ and $M$.  
As all experiments in this paper are performed with the symmetric top molecule ${\rm CH_3F}$, we summarize the associated molecular constants in table~\ref{tab:const}.

\begin{table}[h]
\centering
 \begin{tabular}{ l| c| r }

  constant & symbol & value  \\ \hline \hline
  dipole moment & $\mu$ & \unit[1.85]{D} \\ \hline
  rotational constants & $A_0$ & \unit[155.352]{GHz} \\ \hline
    & $B_0$ & \unit[25.536]{GHz} \\ \hline
   centrifugal distortion & $D_J$ & \unit[60.217]{kHz} \\ \hline
   & $D_{JK}$ & \unit[439.57]{kHz} \\ \hline
   & $D_{K}$ & \unit[2106.923]{kHz} \\ \hline
\end{tabular}
\caption{Summary of molecular constants for CH$_3$F~\cite{Huttner2010}.}
\label{tab:const}
\end{table}

In addition to rotational states, molecules can populate excited vibrational states in various vibrational modes. 
In this paper we consider molecules that almost exclusively populate the vibrational ground state at room temperature. However, we are able to deliberately drive a vibrational transition and use the subsequent spontaneous decay for state manipulation. All state manipulation, rotational and vibrational, is carried out by driving electric dipole transitions and exploiting spontaneous decays that obey the selection rules $\Delta J,\Delta M=\pm1,0$ and $\Delta K=0$.

\subsection{Microwave depletion}
\label{sec:Scheme.MWD}

\begin{figure}[t]
\centering
\includegraphics{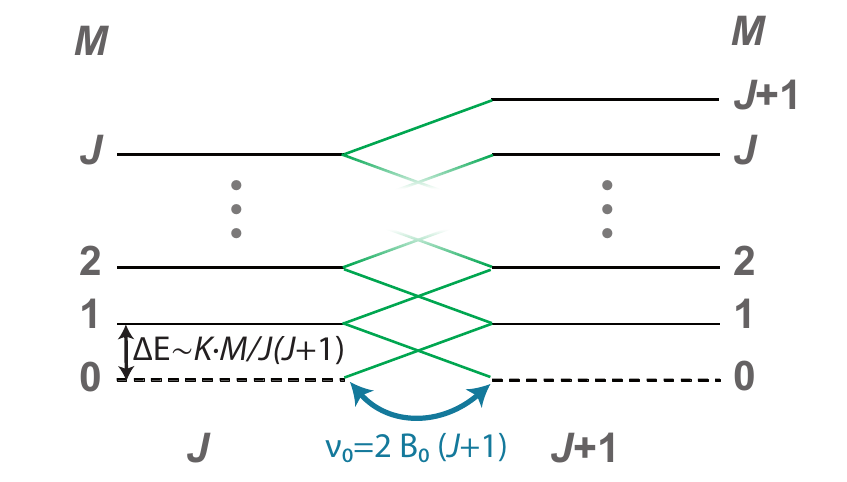}
\caption{Level scheme for microwave depletion (MWD). Untrapped negative $M$-sublevels are omitted. Coupling all $|J;  K;  M\rangle \leftrightarrow |J+1;  K;  M\pm 1\rangle$ transitions (fixed $J$ and $K$) with microwaves (green lines) transfers all molecules populating these states to the untrapped $M=0$ states. Note that in addition to the Stark splitting between neighboring $M$ shown in the figure, a much larger offset energy of about $E=2hB_0 (J+1)$ exists between the states $J$ and $J+1$, as indicated by the bent arrow.}
\label{fig:schemeMWD}
\end{figure}

The transition frequency between rotational states depends on the quantum numbers $J$ and to some extend on $K$ and $M$ (see above). We take advantage of this for our first state detection method, microwave depletion (MWD). By driving microwave transitions between neighboring rotational $J$ states we selectively detect sets of rotational states. As shown in figure~\ref{fig:schemeMWD}, we drive $|J;  K;  M\rangle \leftrightarrow |J+1;  K;  M\pm 1\rangle$ transitions for a specific $J$ and $K$ while addressing all $M$. This couples all states of our set to the untrapped $M=0$ states, successively removing these states from the trapped ensemble. 
 
MWD as just described can be used to detect the population of sets of rotational states with quantum numbers $J$ and $J+1$. Detecting the population of a single rotational $J$ state however is not directly possible as the $J$ state must be coupled to the neighboring state with $J+1$ and/or $J-1$. This problem can in general be solved by performing two measurements and taking the difference. For example, a first measurement gives the population of states with $J,J+1$ and $J+2$ and a second the population of states with $J+1$ and $J+2$. The difference of both measurements yields the population of a single rotational state with the quantum number $J$. 

\subsection{Infrared depletion}
\label{sec:Scheme.IRD}

\begin{figure}[t]
\centering
\includegraphics{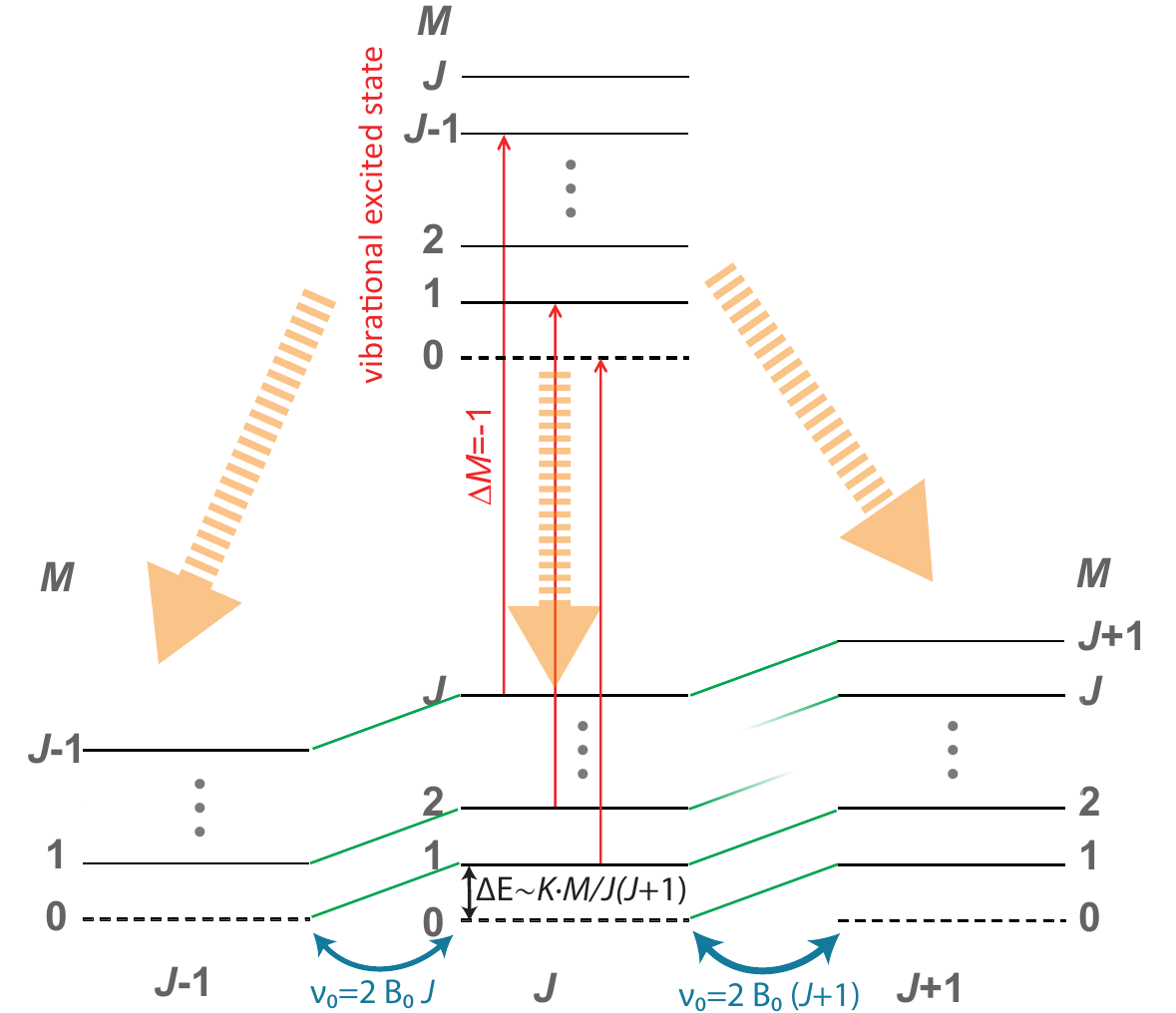}
\caption{Level scheme for infrared depletion (IRD). The laser addresses the first vibrational excited state of a parallel vibrational mode while driving a $\Delta J=0$, $\Delta M=-1$ transition (red arrows). Since the Stark splitting is almost the same in the excited state as in the ground state, we address all $M$-substates simultaneously. The spontaneous decay (big orange arrows) leads to a population transfer to lower lying $M$ states. In the process the $J\pm1$ states get populated, which are coupled to the $J$ state using microwaves (green lines). In combination, the entire population of the shown set of states is transferred to the untrapped $M=0$ states. Note that in the case of $|K|=J$, no $J-1$ states exists, as $K\le J$.} 
\label{fig:schemeIRD}
\end{figure}

Our second depletion method uses optical pumping via a vibrational excitation to deplete the population of the addressed states. These vibrational transitions typically lie in the infrared and we hence entitle this method infrared depletion (IRD). The scheme is shown in figure~\ref{fig:schemeIRD}. A $\Delta M=-1$ Q-branch vibrational transition is driven to pump the population to lower lying $M$-substates thus  successively transferring the population to the untrapped $M=0$ states. In addition, a spontaneous decay can change $J$ by $\pm 1$ leading to population of the states $J-1$ and $J+1$. As a detection via depletion is only useful if the population of the chosen set of states is entirely transferred to untrapped states we in addition have to deplete the states with $J-1$ and $J+1$. This can be achieved with appropriate microwave couplings as shown in figure~\ref{fig:schemeIRD}. Note that in the case of $|K|=J$, spontaneous decay to a state with $J-1$ is forbidden as $K$ is conserved and has to be equal to or smaller than $J$. Thus, only states with the two quantum numbers $J$ and $J+1$ are depleted.   

It is needless to say that using P- and R-branch vibrational transitions for IRD is a valid alternative, especially as it could eliminate the need for coupling the $J$ and $J\pm1$ states with microwaves. However, the frequencies for P- and R-branch transition are spread over a large range. In the Q-branch, the close proximity of the transitions from all rotational states has the big advantage that several rotational states can be addressed with the same laser source (see later discussion). 

\subsection{General comparison of MWD and IRD}
\label{sec:Schemes.Comp}
Here we summarize some general aspects of MWD and IRD for a comparison. First, the timescale for depletion has distinct reasons for MWD and IRD. Whereas the available microwave power sets the timescale for microwave depletion, for IRD it is given by the spontaneous decay rate of the vibrational excited state.

Second, the distribution of lines in the MW and IR spectrum is determined by different regularities. As we will see in the course of this paper, this can be exploited to master the main challenge of our detection: depleting the states of interest while avoiding to address unwanted states with close lying transition frequencies. The frequency for driving rotational transitions between $J$ and $J+1$ states is in first approximation given by $\nu_0=2B_0 (J+1)$ (see equation~\ref{eq:rot}), and hence differs by at least $2B_0$ for different $J$. A dependency on $K$, however, is only introduced by small corrections due to centrifugal distortion and the Stark shift of the $M$-sublevels, where both shifts are much smaller than $B_0$. Resolving the $J$ quantum number with microwaves is thus substantially easier than discriminating different $K$ or $M$.
Q-branch IR transitions on the other hand are more randomly distributed because the rotational constants of the excited states can be quite different for different vibrational modes. Couplings between vibrational modes can additionally shift the transition frequencies. Thus each vibrational transition has to be investigated individually with respect to resolving $J$ and $K$. For CH$_3$F we investigate the transitions of interest in section~\ref{sec:Sel}.

A final difference is that IRD employs fewer MW frequencies than MWD (see figure~\ref{fig:schemeIRD}). Moreover, the transitions needed for IRD have in general higher Clebsch Gordan coefficients and smaller differential Stark shifts than the transitions additionally needed for MWD. Thus less overall MW power is needed for IRD hence reducing the probability to drive unwanted transitions. As we will see, this is an advantage of IRD. 

\subsection{Single rotational $M$-substate detection}
\label{sec:Scheme.M}

Up to now we have discussed the depletion of sets of states while addressing all trapped $M$-substates. We now consider the detection of the population of a single $M$-substate $|J;K;M\rangle$. While a depletion of a single $M$-substate alone is only possible for $M=1$, for $M>1$ it is at least theoretically possible to deplete two sets of rotational states which differ by the single $M$-substate of interest. 

As an example, we discuss the depletion of the single $M$-substate $|J+1;  K;  M=J+1\rangle$ using MWD and IRD (see figures~\ref{fig:schemeMWD} and~\ref{fig:schemeIRD}). In both cases the MW coupling $|J;  K;  M=J\rangle \leftrightarrow |J+1;  K;  M=J+1\rangle$ is the only one that addresses the $|J+1;  K;  M=J+1\rangle$ state. The population of this state is detected by running the whole experimental sequence twice: The depletion (MWD or IRD) is applied once with driving the $|J;  K;  M=J\rangle \leftrightarrow |J+1;  K;  M=J+1\rangle$ transition and once without. The difference of both measurements yields the population in the $|J+1;  K;  M=J+1\rangle$ state.

An extension to other single $M$-substates is possible but at the same time more challenging due to experimental limitations. Using MWD any single $M$-substate can be detected by coupling all $M$-substates up to the desired single $M$-substate and measuring twice as explained above. To be able to implement this scheme, a sufficient spectral resolution is needed to resolve all individual transitions. In the case of IRD one has to keep in mind that spontaneous decay can lead to population transfer to the state of interest.

\section{Experimental setup and sequence}
\label{sec:Exp}

\begin{figure}[t]
\centering
\includegraphics{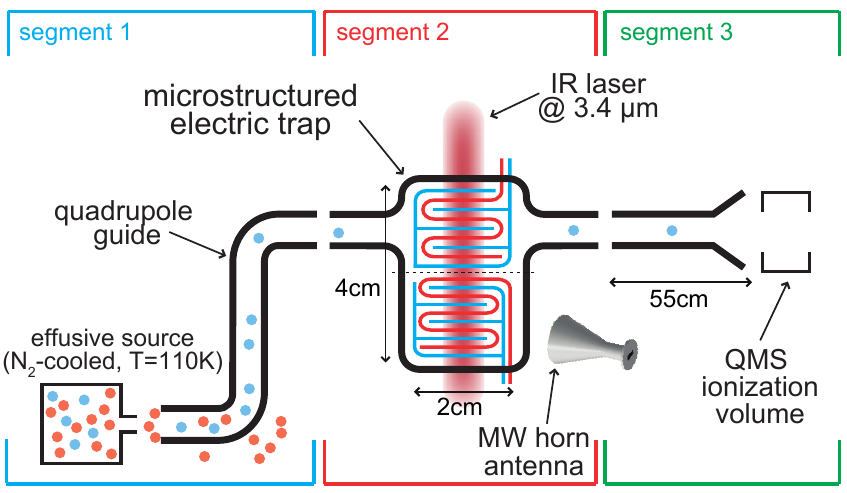}
\caption{Experimental setup. Molecules are loaded from a thermal liquid-nitrogen cooled source to the trap via an electric quadrupole guide. For detection, the molecules are guided to a quadrupole mass spectrometer. See reference~\cite{Englert2011} for more details.}
\label{fig:setup}
\end{figure}

\label{sec:Exp.Scheme}

\begin{figure}[t]
\centering
\includegraphics[width=1\textwidth]{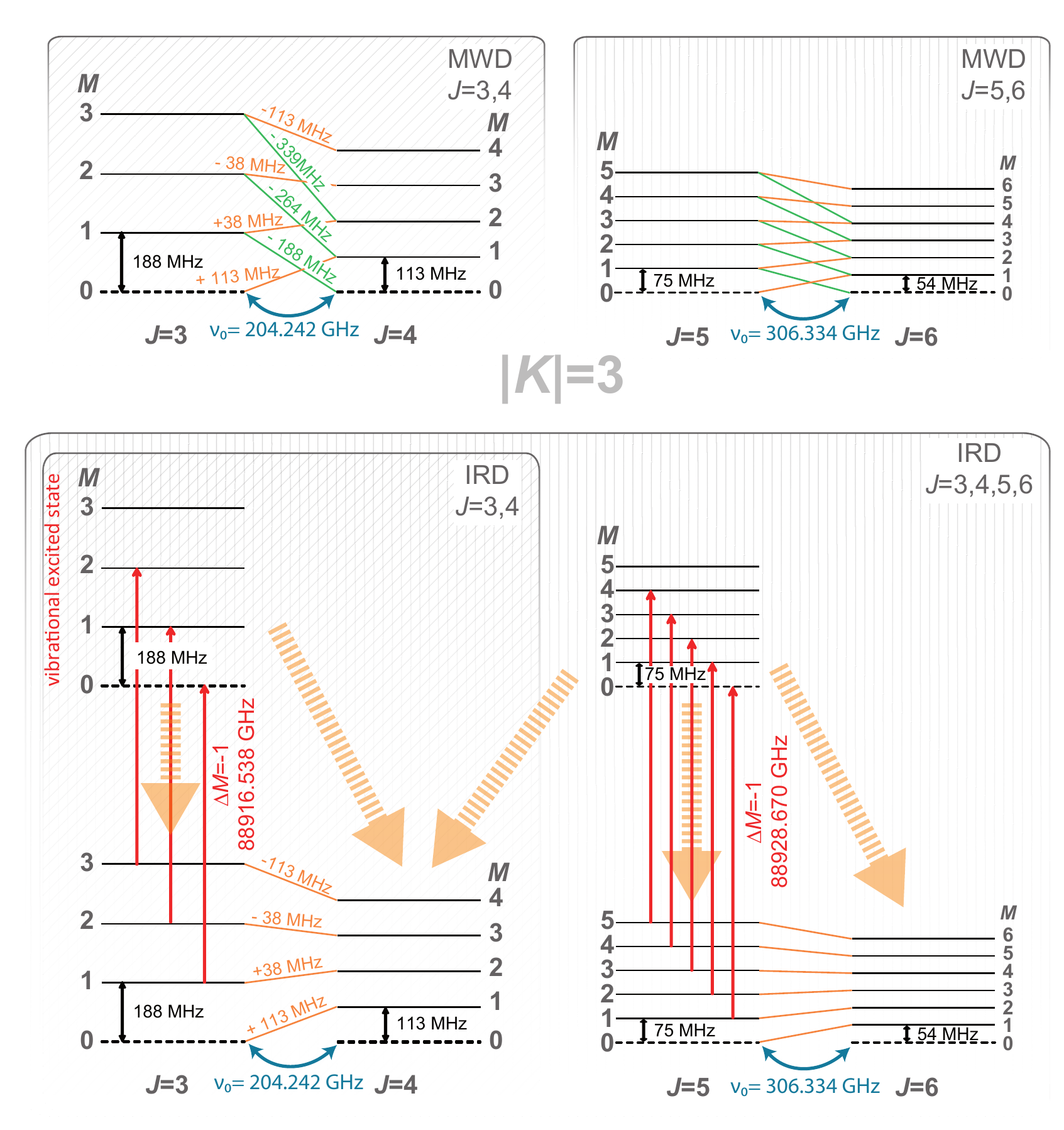}
\caption{Experimentally implemented depletion schemes for the rotational states $|3,4,5,6;  3;  M\rangle$ for MWD and IRD. Stark shifts are plotted to scale and all involved transition frequencies are calculated for the typical electric field in the homogeneous field region of our trap ($\mathcal{E}=\unit[815]{V/cm}$ with $V=\pm\unit[1.8]{kV}$ applied to the microstructures and an offset of $V=\pm\unit[90]{V}$ applied between the capacitor plates; see~\cite{Englert2011} for details). $\nu_0$ is the MW transition frequency between the neighboring $J$ states for zero electric field (bent blue arrows). The applied MW transitions between the $M$-substates are marked in green and orange and the corresponding frequencies specify the detuning from $\nu_0$. The big orange arrows indicate the spontaneous decay channels.}
\label{fig:schemes}
\end{figure}

The experimental setup in which rotational state detection has been implemented is shown in figure~\ref{fig:schemeMWD}. Here, the central part is our homogeneous-field electric trap. A unique design based on trapping molecules between a pair of microstructured capacitor plates allows a tunable homogeneous electric field to be applied in a large fraction of the trap volume~\cite{Englert2011,Zeppenfeld2013,Zeppenfeld2009}. Homogeneous fields are essential since Stark broadening due to the trapping fields is the limiting factor in resolving different rotational transitions inside our trap.

\subsection{Experimental sequence}

For all measurements discussed in this paper the underlying experimental sequences are similar: Initially, molecules generated by velocity filtering via a quadrupole electric guide~\cite{Junglen2004} are loaded into the electric trap for \unit[16]{s}. Subsequently, molecules are stored in the trap, allowing a depletion measurement to be performed. Finally, during the last \unit[12]{s} molecules are unloaded and guided to a quadrupole mass spectrometer for detection. Sisyphus cooling can be optionally applied to the molecules in the states $|3, 4;  3;  M\rangle$ during loading and during the initial period of storage~\cite{Zeppenfeld2012a}. In particular, this substantially increases the fraction of molecules in the states $|3, 4;  3;  M\rangle$, which is useful for a number of measurements in this paper.

\subsection{Experimental implementation of the depletion schemes}

Although we present a general approach to detecting rotational states, we focus on a specific set of states for the experimental demonstration with CH$_3$F. Here, the four lowest $J$ states with $|K|=3$ are well suited as states with $|K|=3$ have an enhanced population in our trap due to spin statistics. Figure~\ref{fig:schemes} shows the detailed depletion schemes experimentally demonstrated in this paper, specifically with the Stark shifts to scale and including all transition frequencies. 

The vibrational excitations of the $v_1$ symmetric CH-strech mode are used for IRD. The IR transitions are driven by a continuous-wave optical parametric oscillator (OPO) that is operated at an idler wavelength of \unit[3.4]{\textmu m}. The frequency is stabilized  by locking the pump beam at \unit[1064]{nm} and the signal at about \unit[1555]{nm} to a frequency comb. A great feature of the OPO system is the ability of mode-hop free tuning by about \unit[80]{GHz} with a piezo control of the pump laser. With a recently developed, fast ramping and locking system, we are able to ramp the frequency within our tuning range and relock to the frequency comb within milliseconds. For example, a ramp between the two needed vibrational transitions from $J=3$ and $J=5$ for depletion, separated \unit[12]{GHz}, can be performed in \unit[7]{ms}. This enables us to drive the two vibrational transitions quasi-simultaneously by changing frequencies every \unit[20]{ms}---fast compared to the decay rate of \unit[15]{Hz} of the vibrational excited state.

The microwaves at \unit[200]{GHz} for coupling the $J=3$ and $J=4$ states and at \unit[300]{GHz} for coupling the $J=5$ and $J=6$ states are produced with amplifier-multiplier chains. An Aeroflex frequency synthesizer  producing up to \unit[18]{GHz} serves as a source. Due to the ability to switch the output frequency within a few \unit[]{\textmu s} all needed MW transitions can be driven quasi-simultaneously. We cycle all frequencies with a rate that is fast compared to the total rate with which we drive rotational transitions. We tune the duty cycle of the individual MW frequencies to adjust their effective power.

\section{Spectral resolution inside our homogeneous-field electric trap}
\label{sec:Limit}

\begin{figure}[t]
\centering
\includegraphics{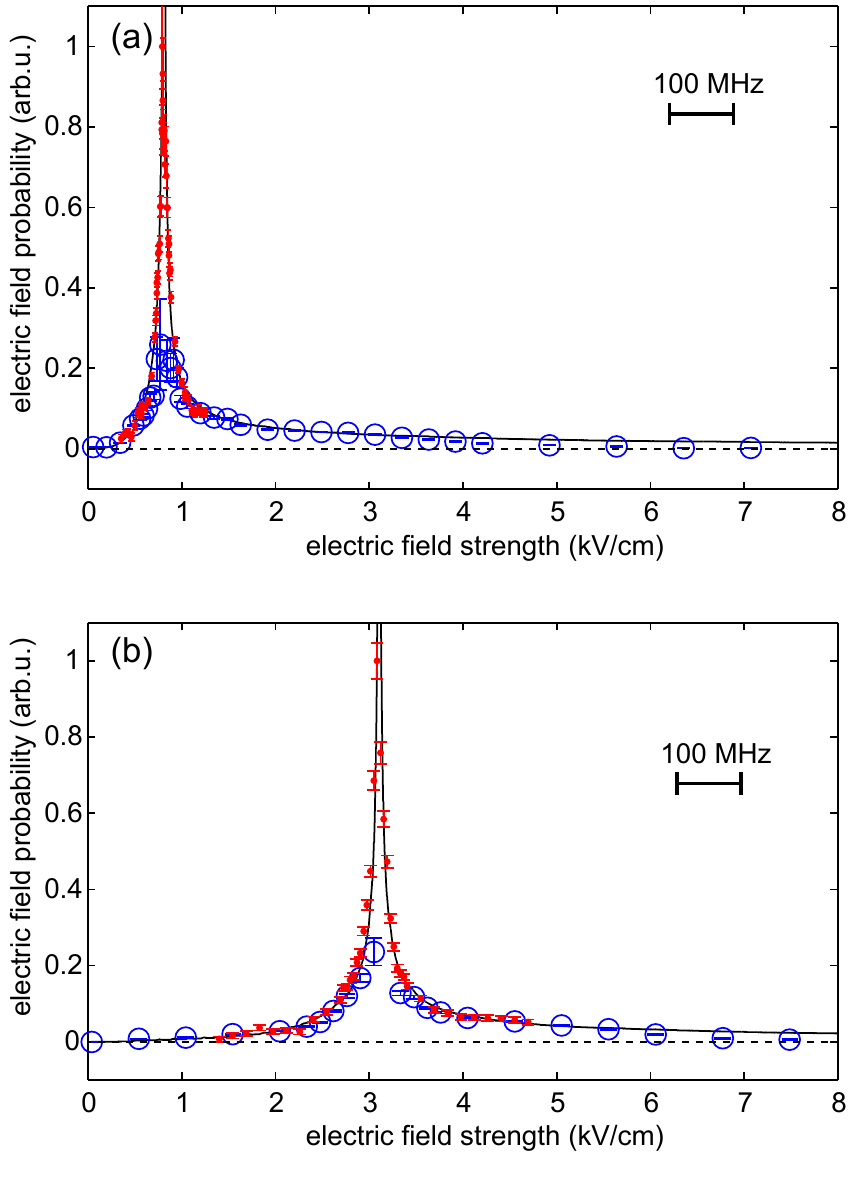}
\caption{Measured and simulated (black line) electric field distribution in the electric trap. We measured the spectral line shape of the single MW transition $|3;  3;  3\rangle \leftrightarrow |4;  3;  4\rangle$ to obtain the electric field distribution. For comparison with the simulation the measured depletion signal is converted into the electric field distribution as explained in the main text. The frequency scale for the unconverted data is indicated. To resolve the flanks, we measured with a higher MW power (blue circles). In addition, we performed a narrow scan at lower power around the peak value (red dots). Part~(a) is measured with the electric fields typically used in our experiments: $\pm \unit[1.8]{kV}$ applied at the microstructures and an offset of $\pm\unit[90]{V}$ between the capacitor plates. Part~(b) is measured with an increased offset of $\pm\unit[450]{V}$ between the plates.}
\label{fig:FieldDist}
\end{figure}

The main ingredient of our experiment is our homogeneous-field electric trap. This trap allows us to dramatically reduce the problem of Stark broadening and thus significantly increases the spectral resolution compared to, e.g., quadrupole traps. As the spectral resolution is important for achieving state selectivity, we carefully investigate our electric field distribution. Specifically, we measure the spectrum of the single MW transition $|3;  3;  3\rangle \leftrightarrow |4;  3;  4\rangle$, where the line shape of this transition is almost exclusively given by the Stark broadening and thus reflects the electric field distribution. The extracted electric field distribution can then be used to calculate the line shape of any other transition of interest.

For the measurement we used a scheme based on the single $M$ state detection via infrared depletion to detect the population in the state $|4;  3;  4\rangle$. As explained in section~\ref{sec:Scheme.M}, driving IR transitions from $J=3$ with $\Delta J=0$ and  $\Delta M=-1$ and coupling all $|3;  3;  M\ne3\rangle \leftrightarrow |4;  3;  M+1\rangle$ transitions with microwaves leads to depletion of all states within the $|3,4;  3;  M\rangle$ manifold except the $|4;  3;  4\rangle$ state. We verified that the $|3;  3;  M\ne3\rangle \leftrightarrow |4;  3;  M+1\rangle$ transitions are saturated. Thus, by adding a fourth MW frequency, any additional depletion can only be caused by the depletion of the $|4;  3;  4\rangle$ state via our target transition $|3;  3;  3\rangle \leftrightarrow |4;  3;  4\rangle$. To obtain the line shape of the transition we varied this MW frequency. The $|3;  3;  3\rangle \leftrightarrow |4;  3;  4\rangle$ transition is always driven resonantly at the corresponding electric field values. We thereby probe the probability for a molecule to be at a position in the trap with the given electric field strength. This probability is essentially the electric field distribution.  

To improve statistics the measurement is performed with a molecular ensemble cooled to \unit[150]{mK} (see section~\ref{sec:Exp}). Before unloading we applied \unit[2]{s} of depletion as explained above. As the occurrence probability of the electric field values varies by more than an order of magnitude, we perform two MW frequency scans. First, we use a higher power of the scanning MW that is adjusted to measure the flanks of our line shape but leads to a complete depletion at the peak. Second, we perform a narrow scan around the peak value with about ten times less power.

To extract the electric field distribution from our data we use a simple theoretical model. The rate $\Gamma$ with which molecules are depleted from the $|4;  3;  4\rangle$ state using a specific MW frequency is proportional to the probability $\rho(\mathcal{E})$ to find an electric field strength $\mathcal{E}$ inside the trap at which the transition frequency is resonant to the applied MW frequency, i.e., $\Gamma \propto \rho(\mathcal{E})$. 
Based on the expectation that the number of molecules $N$ remaining in the state $|4;  3;  4\rangle$ is given by $N \propto \exp(-\Gamma \cdot T)$ with $T$ being the depletion time, our electric field distribution can be calculated via $\rho(\mathcal{E})\propto \log(N)$. 
We verified that the dependence of $N$ on $T$ can indeed be well approximated with an exponential, validating the theoretical approach.

The result of the measurement and the simulation are shown in figure~\ref{fig:FieldDist} for two different electric field configurations. Our measurements have been converted into the electric field distribution using the theoretical approach described above. The result fits nicely to the simulation where only a small horizontal scaling by less than 5\,\%, attributed to inaccuracies in the simulation, is needed to overlap the two. The peak position is mainly given by the offset field due to the voltages applied to the capacitor plates. We measured (a) with a typical offset voltage between the capacitor plates of $\pm\unit[90]{V}$ and (b) with $\pm\unit[450]{V}$. The resulting FWHM is roughly \unit[100]{V/cm} at a peak position of \unit[815]{V/cm} in (a) and \unit[120]{V/cm} at \unit[3.11]{kV/cm} in (b), corresponding to a relative width of 12\,\% and 4\,\%, respectively. This shows that the electric field is homogeneous in a large fraction of the trap volume. 

In addition to the low homogeneous field also higher electric fields for trapping are present in the trap. These trapping fields lead to the long tail of the distribution. This long tail has consequences for the ability to resolve single transitions in any of our experiments. In particular, transitions which would be separated by hundreds of MHz in the homogeneous field region are Stark shifted into resonance. 
Even though the driving rate of such transitions is substantially supressed, the residual driving can still cause unwanted effects. An analysis of the consequences based on rates is provided in section~\ref{sec:Rates}.

The excellent agreement of the measurement and the simulation shows that our electric field distribution is well understood. In particular it proves that we indeed drive the single MW transition $|3;  3;  3\rangle \leftrightarrow |4;  3;  4\rangle$ as shape and peak position of the spectrum scale with trap voltages as expected.  Note that in figure~\ref{fig:FieldDist} the measured distribution lies below the simulation at higher fields. This cutoff is due to the limited kinetic energy of the molecules preventing them from reaching the high-field regions.

\section{$J$,$K$ and $M$ selectivity of IR and MW transitions}
\label{sec:Sel}

\begin{figure}[t]
\centering
\includegraphics{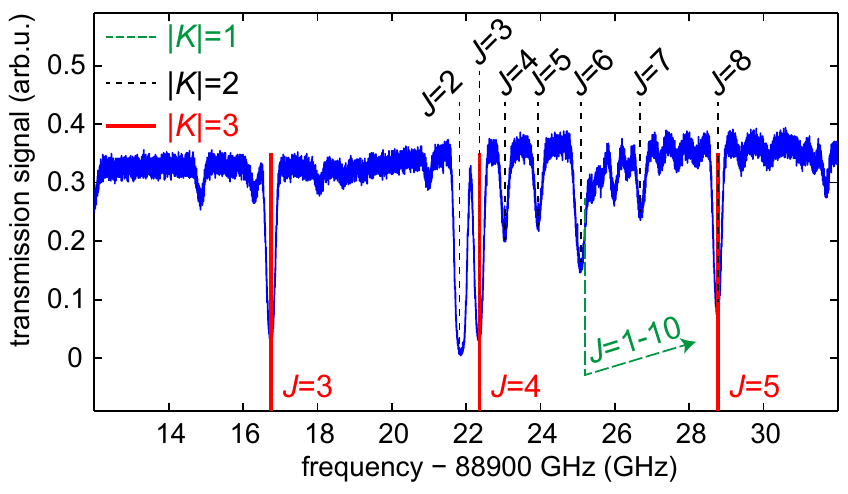}
\caption{Doppler-broadened absorption spectrum of the $v_1$ vibrational mode of CH$_3$F obtained with a room temperature cell. We show the measurement in the vicinity of the relevant transitions. The transition frequencies for $|K|=1$ lie quite close together and increase for higher $J$, as indicated by the arrow. We verified that all transitions in the Q-branch with $|K|=4$ to $12$ lie well outside the shown frequency range~\cite{Graner1981}. All non-identified lines thus correspond either to higher $J,K$ states or to other vibrational modes and can be neglected for our depletion schemes. The transitions from $J=3$ and $J=5$ for $|K|=3$ are well isolated, with only the transitions from states with $J>8$, $|K|=1$ lying in the vicinity of the transition from $J=5$, $|K|=3$.}
\label{fig:IRSpectrum}
\end{figure}

\begin{figure}[t]
\centering
\includegraphics{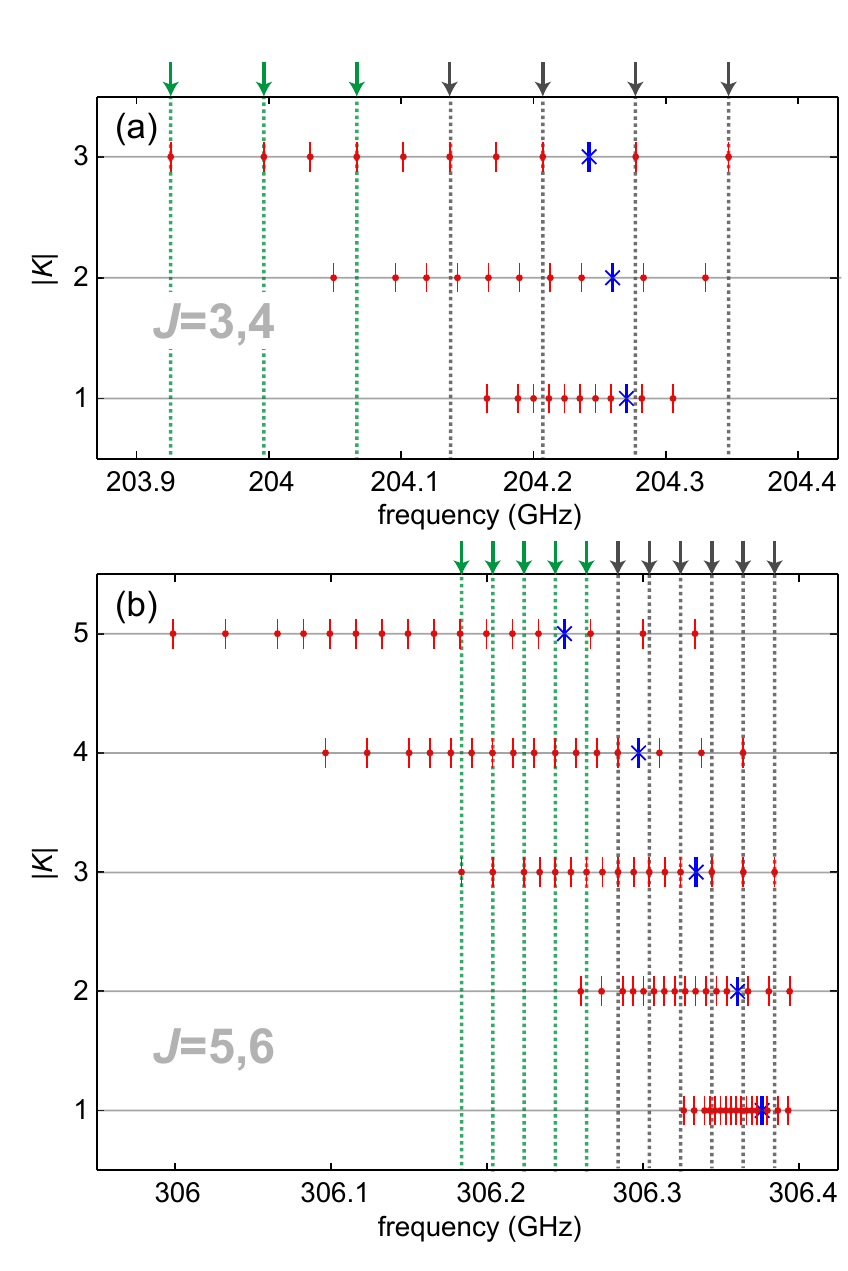}
\caption{Rotational transition frequencies for different $K$-values for the $J=3\leftrightarrow J=4$ MW transitions~(a) and the $J=5 \leftrightarrow J=6$ MW transitions~(b). All transition frequencies are calculated for the typical homogeneous electric field in the trap. The blue crosses indicate the frequencies for zero electric field for the individual $K$ states and the red dots represent the frequencies of the Stark-shifted dipole-allowed transitions between $M$-substates. The arrows at the top mark the transition frequencies needed for MWD in the $|K|=3$ manifold and the dashed lines are guides to the eye. For IRD only the highest four (a) or six (b) frequencies are needed. For an analysis of the $J,K$ and $M$ selectivity the frequency separations have to be compared to the spectral resolution obtained from the measurement in figure~\ref{fig:FieldDist}(a).}
\label{fig:Kvalues}
\end{figure}

Following the previous analysis of our spectral resolution, we now discuss the $J$,$K$ and $M$ selectivity of our depletion techniques. For this purpose we have to consider whether there are other transitions close to the ones that we want to drive. We then have to compare the frequency separations to our spectral resolution. General considerations were given in section~\ref{sec:Schemes.Comp}. Here, we present a detailed analysis for the MW transitions $|3;  3;  M\rangle \leftrightarrow |4;  3;  M\pm1\rangle$ and $|5;  3;  M\rangle \leftrightarrow |6;  3;  M\pm1\rangle$ as well as for the IR Q-branch transitions from the $J=3$ and $J=5$ states used for our depletion schemes as shown in figure~\ref{fig:schemes}.

IR transition frequencies of the $v_1$ vibrational mode are provided in~\cite{Graner1981} and were verified via saturated absorption spectroscopy in a room temperature cell. The transition frequencies were determined with sub-\unit[]{MHz} resolution using a frequency comb. We found that luckily the Q-branch IR transition from the $J=3,|K|=3$ state is well isolated, with the closest relevant transition from a different state being several GHz away (see figure~\ref{fig:IRSpectrum}). Our spectral resolution is clearly better than this and thus the driving of this transition is $K$ and $J$ selective.
The IR transition from the $J=5,|K|=3$ state is less isolated. Here, within several GHz we find also the IR transitions from the $J=7,8,|K|=2$ states and $J=8,9,10,|K|=1$ states. However, inside our trap, these states are in sum populated by at most a few percent and can therefore be neglected.  Despite the fact that both transitions can be regarded as being perfectly $J$ and $K$ selective, resolving $M$ is not possible while using the Q-branch. The Stark shifts of the vibrational excited states are almost the same as in the ground state and therefore the Stark shift does not lead to a separation of transitions with different $M$ quantum numbers. 

The MW transition frequencies between neighboring rotational $J$-states differ by at least \unit[50]{GHz} for different $J$. Thus an unintended addressing of states with other $J$ can be excluded, and driving MW transitions can be considered as being perfectly $J$ selective. To find out whether the driving of MW transitions can also resolve the $K$ and $M$ quantum numbers, we calculate all transition frequencies between the rotational states $|3;  K ;  M \rangle \leftrightarrow | 4 ;  K ;  M\pm 1\rangle$ according to equations~\ref{eq:rot} and~\ref{eq:StarkShift}, using the typical electric field value of the homogeneous field region in our trap. The result is shown in figure~\ref{fig:Kvalues}(a). A comparison of the frequency differences and our electric field distribution (section~\ref{sec:Limit}) shows that it is not possible to fully resolve the $K$ or the $M$ quantum numbers. The rate at which the non-addressed transitions are driven and the consequences for the state selectivity will be analyzed in the next section.

We similarly calculated the transition frequencies between the rotational states $|5;  K ;  M \rangle \leftrightarrow | 6 ;  K ;  M\pm 1\rangle$ (figure~\ref{fig:Kvalues}(b)). 
Here at least some of the transitions between states with $|K|> 3$ are better separated from the frequencies of the $|K|=3$ transitions. We therefore expect that states with $|K|> 3$ are less affected than states with $|K|< 3$ when addressing the states $J=5,6$ with $|K|= 3$.

\section{Rate model for the dynamics of the depletion}
\label{sec:Rates}

To quantify the dynamics of our depletion techniques we have set up rate models. These rate models allow us to analyze the timescales of depletion of the states of interest with $|K|=3$ and of unwanted states with $|K|\ne3$. In addition we include population transfer caused by blackbody radiation and discuss its effect on the depletion signal.

For our entire analysis we ignore coherent processes and study the population transfer using the rate equations. This assumption can be validated by the experimental parameters: The MW rate is at most on the order of hundreds of \unit[]{Hz} and has to be compared to the Stark broadening in the central trap region of many \unit[]{MHz}. The argument for the infrared radiation is slightly different. Although the laser intensity might locally be high enough to drive coherent processes, the molecules pass the small laser beam only once in a while in our large trapping volume. 

\subsection{Rate model}

The rate equations have the usual form
\begin{equation}
\dot{p}_i(t)=\sum_{j}\Gamma_{i,j}p_j(t)
\end{equation}
where $p_i$ is the population in state $i$ and $\Gamma_{i,j}$ is the rate for driving a transition from a state $i$ to a state $j$. We have to take several contributions into account: rotational states coupled with MW, vibrational transitions driven via the laser, the spontaneous decay from the vibrational excited states, and blackbody induced driving.  

Unlike for the spontaneous decay and the blackbody induced population transfer the driving of the MW and IR transitions depends on the applied power. The appropriate rates for the model are determined as follows: For IR transitions, the driving rate is much faster than any other relevant processes in the trap and in particular it is much faster than the spontaneous decay rate of \unit[15]{Hz}. The exact rate hence does not influence the end result as long as it is sufficiently large, and we chose a value of \unit[1]{kHz}.

Two components influence the rate $\Gamma^{MW}_{i,j}$ with which a rotational transition from state $i$ to state $j$ is driven: The wanted driving of this transition in the homogeneous field region of the trap (if applicable) and the unwanted driving of this transition in the inhomogeneous electric field regions due to the other applied MW frequencies needed for MWD or IRD. For both processes the rate can be calculated
 with the following assumptions. We consider applying a single fixed MW frequency $\nu$, which does not necessarily match the transition frequency from state $i$ to state $j$ in the homogeneous electric field region. First of all, the rate $\Gamma^{MW}_{i,j}$ is then proportional to the effective driving power $P$ of the applied MW frequency. Second, it is proportional to $c_{i,j}$, the square of the Clebsch Gordan coefficient for transitions between state $i$ and state $j$. Third, the rate $\Gamma^{MW}_{i,j}$ is proportional to the spectral line shape function $\rho_{i,j}(\nu)$ of the given transition. This can be written in terms of a coefficient $d_{i,j}=\left(\frac{K M_j}{J_j(J_j+1)}-\frac{K M_i}{J_i(J_i+1)} \right)$ for the differential Stark shift of the transition and the electric field distribution $\rho(\mathcal{E})$ as follows. With $\nu=\nu_0+\frac{\mathcal{E}_{i,j}\,\mu}{h}\cdot d_{i,j}$ where $\mathcal{E}_{i,j}$ is the electric field where the transition takes place, we have $ \rho_{i,j}(\nu)=\rho(\mathcal{E}_{i,j})\frac{d\mathcal{E}_{i,j}}{d \nu}\propto\rho(\mathcal{E}_{i,j})\frac{1}{d_{i,j}}$. $\nu_0$ denotes the transition frequency in the absence of any electric fields. The rate $\Gamma^{MW}_{i,j}$ for the single MW frequency is thus given by
\begin{equation}
\Gamma^{MW}_{i,j}\propto \rho(\mathcal{E})\cdot \frac{c_{i,j}}{d_{i,j}}\cdot P.
\label{eq:gamma}
\end{equation}
To obtain the total rate with which the rotational transition from state $i$ to state $j$ is driven, the contributions for the various applied MW frequencies have to be summed up.

\begin{figure}[t]
\centering
\includegraphics{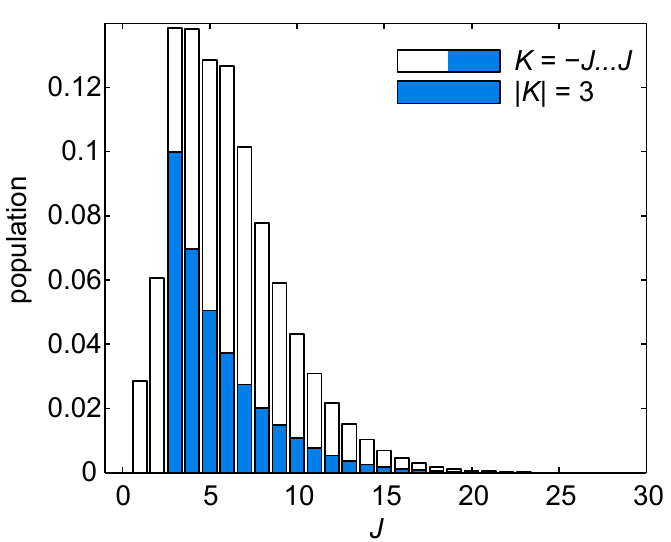}
\caption{Rotational state distribution for CH$_3$F in the electric quadrupole guide for velocity filtering from a thermal source at a temperature of \unit[110]{K}.}
\label{fig:dist}
\end{figure}

In addition to the coupling matrix $\Gamma$, the initial state distribution inside our electric trap is needed. The thermal population of rotational states in our liquid-nitrogen cooled nozzle can be calculated using Maxwell-Boltzmann statistics as no external fields are present. However, the population in the electric quadrupole guide differs as the trapping force depends on the molecular state, and the Maxwell-Boltzmann distribution has to be weighted with the Stark shift squared~\cite{MotschBoosting}. The resulting population distribution is shown in figure~\ref{fig:dist}. For the initial population in the rate models we included the rotational states $J=1$ to $J=8$ with all possible $K$ and $M$ values.

\subsection{Blackbody radiation}
\label{sec:Limit.BB}

Population transfer caused by blackbody radiation influences the dynamics of a state detection via depletion. This population transfer leads to a repopulation of states during the depletion and thus to an enhanced number of depleted molecules. The effect of blackbody radiation can then only be neglected if the timescale for repopulation is much slower than the depletion.

In the past, the effect of blackbody radiation on molecular states was associated with blackbody induced driving of rotational transitions~\cite{Hoekstra2007}. Our molecule, however, has a relatively small rotational constant $B_0$ and thus blackbody induced transitions between rotational states can be neglected. For transitions between vibrational states the transition strengths for various vibrational modes of a molecule can vary over a large range and each vibrational mode has to be considered individually. Indeed, the $v_3$ CF-stretch vibration in $\rm CH_3F$ lies at $\unit[1049]{cm^{-1}}$ and has a spontaneous decay rate of about \unit[13]{Hz}~\cite{Newton1976}. At room temperature this leads to blackbody rate of $\Gamma_{bb}=\unit[0.075]{Hz}$, which is relevant on the timescales of our experiments (see following sections). Note that only dipole allowed  transitions contribute and can therefore only lead to a change of the $M$ and $J$ values with $K$ being conserved.

\subsection{Rate model for MWD}
\label{sec:Rates.MWD}

The experimental implementation of MWD in the states with $J=3,4$, $|K|=3$ involves seven frequencies (see figure~\ref{fig:schemes}), and an efficient depletion can be achieved by driving the corresponding seven transitions with an equal rate. Based on equation~\ref{eq:gamma}, we hence choose the relative power of the MW frequencies according to $P_{i,j}\propto \frac{d_{i,j}}{c_{i,j}}$, which in our experiment is implemented by variation of the duty cycle (see section~\ref{sec:Exp}). For example, to drive the $|3;  3 ;  3\rangle \leftrightarrow | 4 ;  3 ;  2\rangle$ transition with the same rate as the $|3;  3 ;  3\rangle \leftrightarrow | 4 ;  3 ;  4\rangle$ transition, an about 80 times higher MW power is needed. 

Figure~\ref{fig:Rate}(a) shows the result of the rate model for MWD of the rotational states with $J=3,4$, $|K|=3$. Note that the absolute timescale is arbitrary as it purely depends on the total applied MW power. The timescale is chosen to match to the experimental data in section~\ref{sec:Results.Sat}. As can be seen immediately, we deplete states with $|K|=1,2$ and $|K|=3$ on almost the same timescale and it is thus impossible to obtain a $|K|$ dependent MWD depletion signal. 
By including population transfer caused by blackbody radiation (section~\ref{sec:Limit.BB}) we observe an increase of the population with $M=0$ (dashed black curve) on long timescales. 
We repeated the analysis for MWD of the $J=5,6$, $|K|=3$ states and the result is given in figure~\ref{fig:Rate}(b). Again states with $|K|<3$ are depleted on the same timescale as the $|K|=3$ states. In contrast, the $|K|>3$ states are  less affected due to the larger frequency separation (see figure~\ref{fig:Kvalues}). Including blackbody radiation in the rate model results, as before, in an increase of the population with $M=0$ (dashed black curve) on long timescales.

\subsection{Rate model for IRD}

\begin{figure}[t]
\centering
\includegraphics{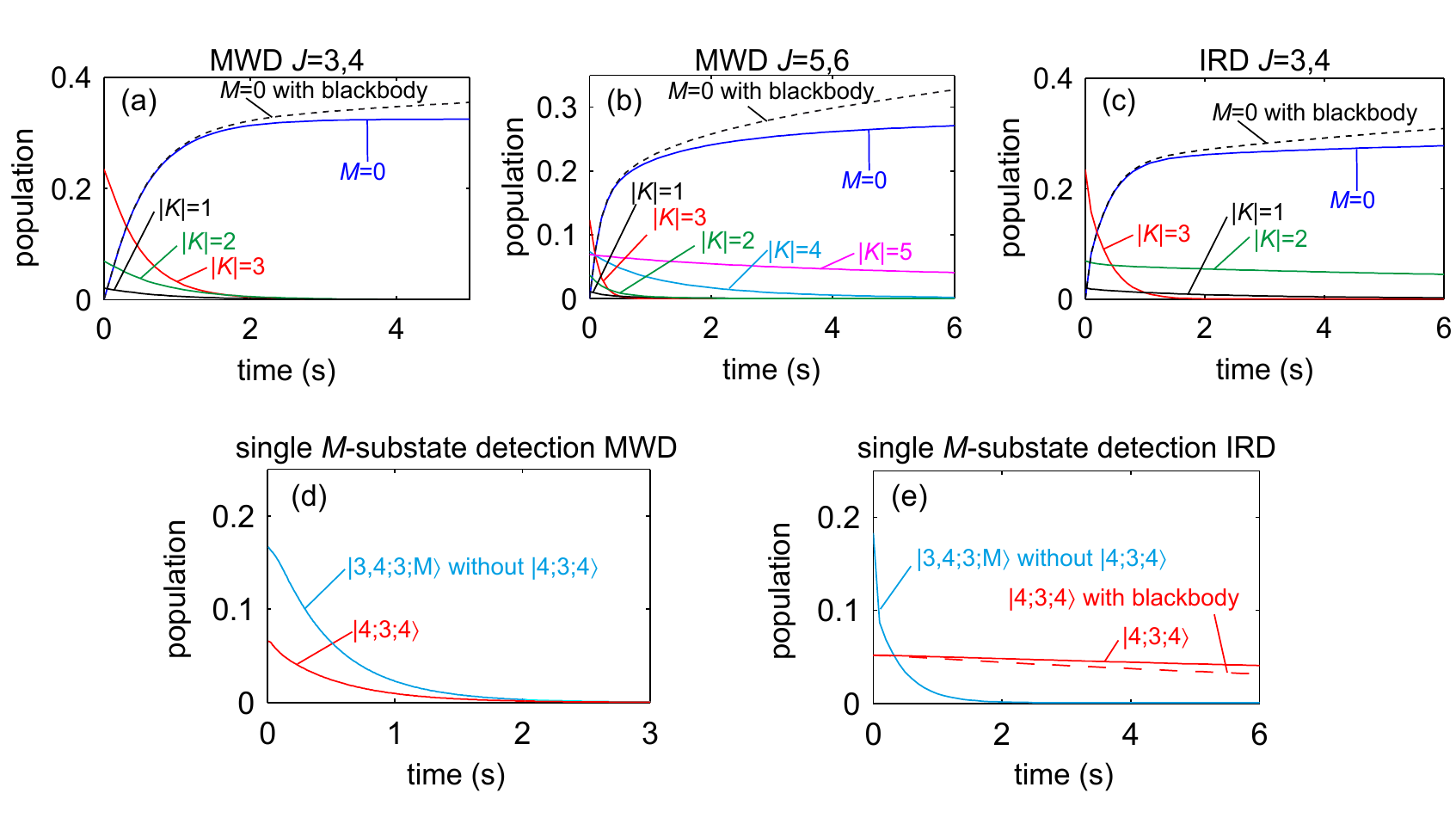}
\caption{
Rate models. (a)-(c) Depletion of sets of states according to the schemes displayed in figure~\ref{fig:schemes}. We show the time evolution of population remaining in trapped states for individual $K$-manifolds (labeled with $|K|=1,2,...$) and of population being transferred to the untrapped $M=0$ states (labeled with $M=0$). (a) MWD for target states $J=3,4$, $|K|=3$. (b) MWD for target states $J=5,6$, $|K|=3$. (c) IRD for target states $J=3,4$, $|K|=3$. The effect of blackbody radiation is indicated by the dashed lines. (d)-(e) Implementation of MWD (d) or IRD (e) without driving the $|3;  3;  3\rangle \leftrightarrow |4;  3;  4\rangle$ transition, used for detection of the population in the single $M$-substate $|4;3;4\rangle$ (see main text). The population remaining in the trap is plotted for the state of interest $|4;3;4\rangle$ and for all other states of the manifold $|3,4;3;M\rangle$.
}
\label{fig:Rate}
\end{figure}

Figure~\ref{fig:Rate}(c) shows the results of the IRD rate model. In contrast to MWD the timescale for depletion is not given by the power of the IR or MW radiation but set by the spontaneous decay rate of the vibrational excited state. After about \unit[2]{s} all molecules of the $J=3,4$, $|K|=3$ manifold are depleted. Compared to MWD, IRD uses only four MW transitions. These transitions have small differential Stark shifts and large Clebsch Gordan coefficients and thus need a relatively low MW power to be driven at a faster rate than the spontaneous decay. Thus states with $|K|=1,2$ are still depleted but on a much slower timescale than for MWD. As a result, the moderate increase of depleted molecules on long timescales is here unlike for MWD not only given by repopulation of the states due to blackbody radiation but also by the slow depletion of states with $|K|\ne3$. 

\subsection{Rate model for single $M$-substate detection}

In section~\ref{sec:Scheme.M} we discussed schemes for a detection of a single $M$ state using a modification of IRD and MWD. We now analyze these schemes for the $|4;  3;  4\rangle$ states using the rate model. Both methods rely on the capability to switch off the driving of the  $|3;  3;  3\rangle \leftrightarrow |4;  3;  4\rangle$ transition. Using the rate models we are now able to test whether it is possible to deplete all molecules within the states described by $J=3,4$, $|K|=3$ apart from the $|4;  3;  4\rangle$ state. 

Figure~\ref{fig:Rate} shows the result of an MWD~(d) and an IRD~(e) simulation where the coupling of the $|3;  3;  3\rangle \leftrightarrow |4;  3;  4\rangle$ transition is not applied. The results presented in (d) show that for MWD the $|4;  3;  4\rangle$ state is depleted at a similar rate as the rest of the $J=3,4$, $|K|=3$ states and thus a single $M$ detection is not possible. This is primarily due to the high power needed to drive the $|3;  3;  3\rangle \leftrightarrow |4;  3;  2\rangle$ transition, as mentioned above.

In contrast, using IRD, the detection of the population of the single $M$ state $|4;  3;  4\rangle$ is possible and was in fact used to measure our electric field distribution in section~\ref{sec:Limit}. The population of the $|4;  3;  4\rangle$ state is still depleted but on a much slower timescale, resulting in an error of the population measurement. Blackbody radiation causes the unwanted loss of population from the single $M$-substate to further increase.

\section{Experimental results}
\label{sec:Results}

In the following we discuss experimental results for our state sensitive detection methods MWD and IRD. First, measurements of the depletion time dependence are presented. These saturation measurements yield the timescale needed for depletion which can then be compared to other timescales such as the population transfer caused by blackbody radiation or the trap lifetime. In addition, we find that the predictions from the rate model (see section~\ref{sec:Rates}) agree well with the measurement results, explaining the effect of blackbody radiation and unwanted depletion of wrong $K$-states. For both problems we discuss solutions in the following section. In particular we show that with a variation of IRD a detection signal can be obtained which is independent of the depletion in wrong $K$-states. 

Knowing the relevant timescales we subsequently investigate the quality of our depletion techniques: We first examine to which extent our methods yield the same result. Second, we prove that all molecules populating the states of interest can be depleted. 

As a final result we measure the population of the $|4;  3;  4\rangle$ state achieving the main goal of our rotational state detection: detecting a single rotational state described by single $J,K$ and $M$ quantum numbers. A time dependent saturation measurement is used to estimate the error of our single $M$-substate detection.   

\subsection{Saturation measurements}
\label{sec:Results.Sat}

All saturation measurements were performed with a similar experimental sequence as described in section~\ref{sec:Exp}. The state sensitive depletion is applied for varying amounts of time after the first second of storage. The total storage time, however, is always the same (\unit[9]{s} for MWD; \unit[7]{s} for IRD) to allow us to ignore trap losses.    

\subsubsection{Microwave depletion}

\label{sec:Results.MWD}
\begin{figure}[t]
\centering
\includegraphics{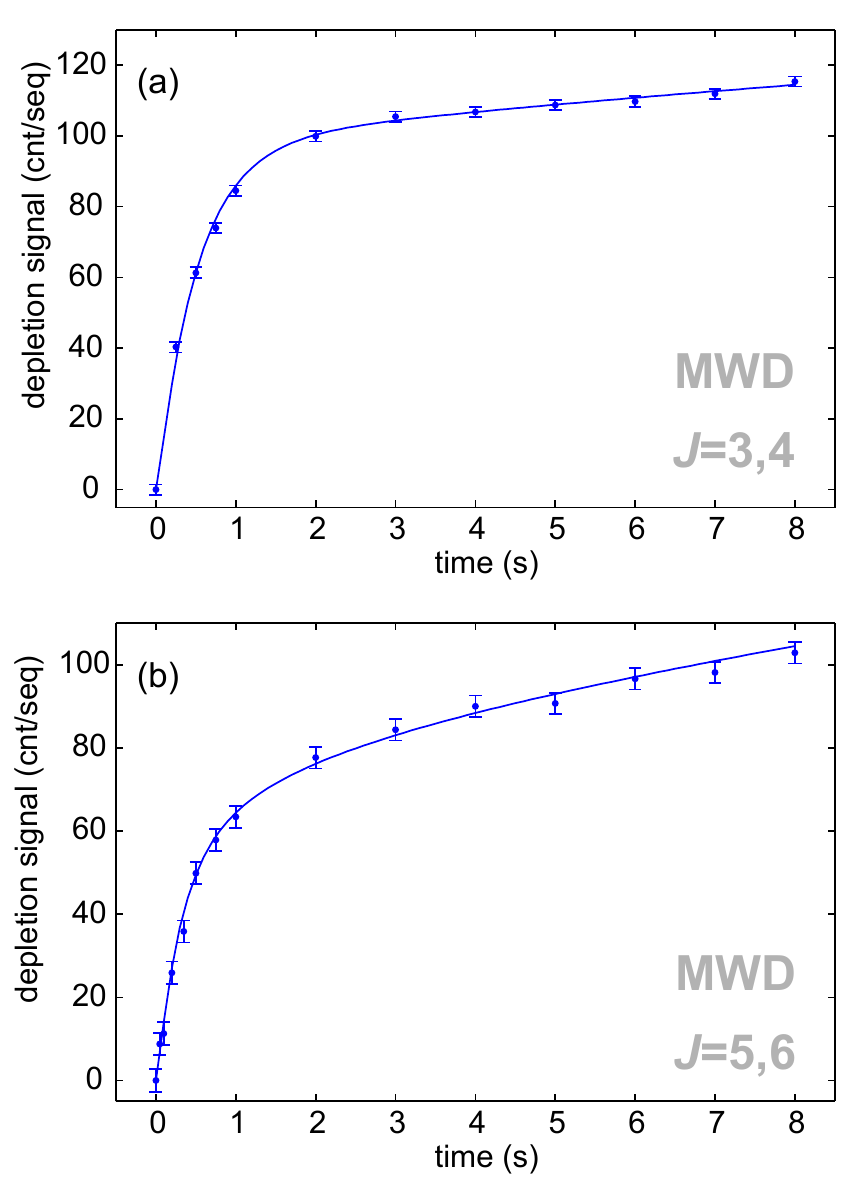}
\caption{ Experimental results and comparison to the rate model for MWD. A saturation measurement of MWD is shown for the $|3,4;  3;  M\rangle$ states in (a) and the $|5,6;  3;  M\rangle$ states in (b). Both measurements show two timescales. The fast increase in (a) is given by the depletion of the entire $J=3,4$ states, including also the unintended depletion of states with $|K|=1,2$. The slow increase results from blackbody induced population transfer, see main text for details. (b) shows the same behavior except for a more pronounced slow increase. In addition to blackbody induced population transfer, this is due to the slow depletion of states with $|K|>3$.}
\label{fig:MWD}
\end{figure}

MWD is implemented according to the scheme discussed in section~\ref{sec:Scheme.MWD} with the experimental details for our rotational states of interest given in figure~\ref{fig:schemes}. The result of the saturation measurement and the rate model for the $J=3,4$, $|K|=3$ states are presented in figure~\ref{fig:MWD}(a) and show an excellent agreement. To fit the rate model to the data we only used two fit parameters: The first one is the total power of the microwaves. This is a single fit parameter as the effective power of the individual applied MW frequencies is set according to $P_{i,j} \propto \frac{d_{i,j}}{c_{i,j}}$ (see section~\ref{sec:Rates.MWD}). The second one is a simple vertical scaling parameter.

The saturation measurement shows two timescales, as expected from the rate model. The fast increase is the depletion of the $J=3,4$ states, where the unwanted $|K|=1,2$ states are depleted almost equally fast as the $|K|=3$ states of interest. A better separation can only be achieved with an even better trap design. Thus signal contributions due to the $|K|=1,2$ states are at the moment only limited by the population of these states. A calculation of the population distribution inside our electric quadrupole guide loaded from a liquid-nitrogen cooled thermal source at \unit[110]{K} shows that 72\,\% of the molecules in the $J=3,4$ manifold populate the $|K|=3$ state (see figure~\ref{fig:dist}). Thus using MWD the measured number of molecules populating states with $J=3,4$ $|K|=3$  is at most overestimated by 40\,\%. 

The slow increase in figure~\ref{fig:MWD}(a) is a result of blackbody induced population transfer. Whereas the timescale for the depletion via MWD is purely limited by the total MW power, the timescale of blackbody induced population transfer is given by the temperature of the setup. Thus the timescales of both processes are independent and a sufficient separation is achievable.

The saturation measurement can now be used to choose the depletion time for state sensitive detection. While measuring the full saturation curve gives additional information, measuring only a single point is sufficient for many purposes and requires substantially fewer measurements. We hereby have to find a compromise: On the one hand, too long depletion times lead to larger errors due to the population transfer caused by blackbody radiation. On the other hand, the depletion of the states of interest should be saturated. We picked \unit[2]{s} for further measurements. Here the influence of blackbody radiation on the detection result is on the order of a few percent and thus almost negligible.

A MWD measurement of the population within the $|5,6;  3;  M\rangle$ state and the corresponding rate model is shown in figure~\ref{fig:MWD}(b). The overall shape of the curve is similar to the one in figure~\ref{fig:MWD}(a) and again agrees nicely with the rate model.  However, compared to figure~\ref{fig:MWD}(a),~(b) shows a much more pronounced increase at longer depletion times. This is because in the manifold of states with $J=5,6$ only about 40\,\% of the molecules populate the states with $|K|=3$. Moreover, as discussed in section~\ref{sec:Rates}, the states with $|K|>3$ contribute on top of the blackbody radiation to the slow increase of the signal. The error for detecting states with $|K|=3$ is here larger than for the $J=3,4$ states, and MWD should only be used to discriminate the total angular momentum $J$. 

\subsubsection{Infrared depletion}

\label{sec:Results.IRD}

\begin{figure}[t]
\centering
\includegraphics{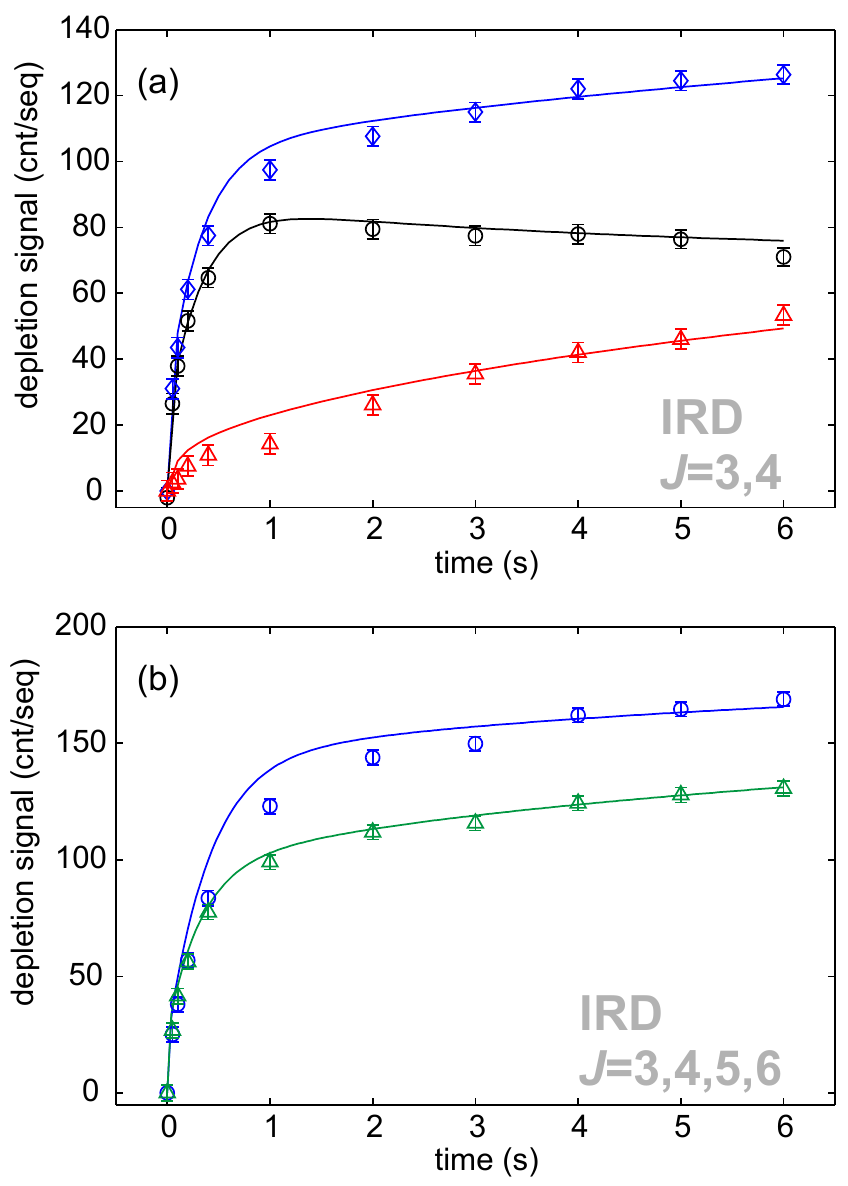}
\caption{Saturation measurement and comparison to the rate model for IRD. (a)~The blue diamonds show the result of an IRD saturation in the states with $J=3,4$, $|K|=3$ according to the scheme shown in figure~\ref{fig:schemes}. The red triangles represent a reference measurement, without applying the laser. The difference of the two (black circles) yields a $|K|$-selective IRD measurement as explained in the main text. (b)~IRD saturation of the states  with $J=3,4,5,6$ and $|K|=3$. The difference of the full IRD signal (blue) and the measurement with the laser only applied to the $J=3$ state (green) yields the $K$-selective signal of molecules populating the states with $J=5,6$ and $|K|=3$.}
\label{fig:IRD}
\end{figure}

The rate model and the experimental result of the saturation measurement for the states with $J=3,4$, $|K|=3$ using IRD is given in figure~\ref{fig:IRD}(a). The implementation is realized according to the scheme discussed in section~\ref{sec:Scheme.IRD} with experimental details given in figure~\ref{fig:schemes}. This leads to the blue curve (blue diamonds) which is in good agreement with our rate model. We denote this method full IRD in contrast to $K$-selective IRD described below. 

As already discussed above, the driving of the infrared transitions is $J$ and $K$ selective and the unwanted depletion of states with $|K|=1,2$ is purely caused by the MW radiation. This timescale is however much slower than the depletion of the states with $|K|=3$ (see section~\ref{sec:Rates}). Thus, in contrast to MWD, the unintended depletion of states with $|K|=1,2$ also contributes to the slow increase of the depletion signal, which is of great advantage: By picking again \unit[2]{s} of depletion for further experiments, the detection signal is less influenced by the unintended depletion of states with $|K|=1,2$, thus reducing the error.
 
This error can completely be eliminated by subtracting out the influence of the MW radiation. For that, we examined the depletion effect of the microwaves with a reference measurement. Here the four MW frequencies for IRD were applied but the laser was left off (\ref{fig:IRD}(a) red triangles). As both measurements (the full IRD and reference measurement) are equally affected by the unintended depletion of states with $|K|\ne3$, the difference of both measurements yields a $|K|$-selective measurement (black circles). The drawback of this $K$-selective IRD is a slightly reduced signal as the reference measurement also includes some depletion of states with $|K|=3$.  

Our IRD scheme can easily be extended to incorporate also the $J=5,6$ rotational states as presented in section~\ref{sec:Exp.Scheme}. Figure~\ref{fig:IRD}(b) shows the corresponding rate model and saturation measurement. The full IRD signal for depleting the $|3,4,5,6;  3;  M\rangle$ states is given by the blue curve (blue circles). In addition, we can deduce the population in the states $|5,6;  3;  M\rangle$ by performing a second measurement without driving the vibrational excitation from $J=5$ (green triangles). This leads to the depletion of the $J=3,4$, $|K|=3$ states as in (a). However, the MW couplings of the $J=5,6$, $|K|=3$ states are additionally present. Thus, the effect of driving other $K$ states with microwaves is contained in this measurement, and the difference of both measurements in~(b) gives a $|K|$-selective signal for the molecule number in the states $|5,6;  3;  M\rangle$.

\subsection{Qualitative investigation of the rotational state depletion methods}
\label{sec:Results.Com}
\begin{figure}[t]
\centering
\includegraphics{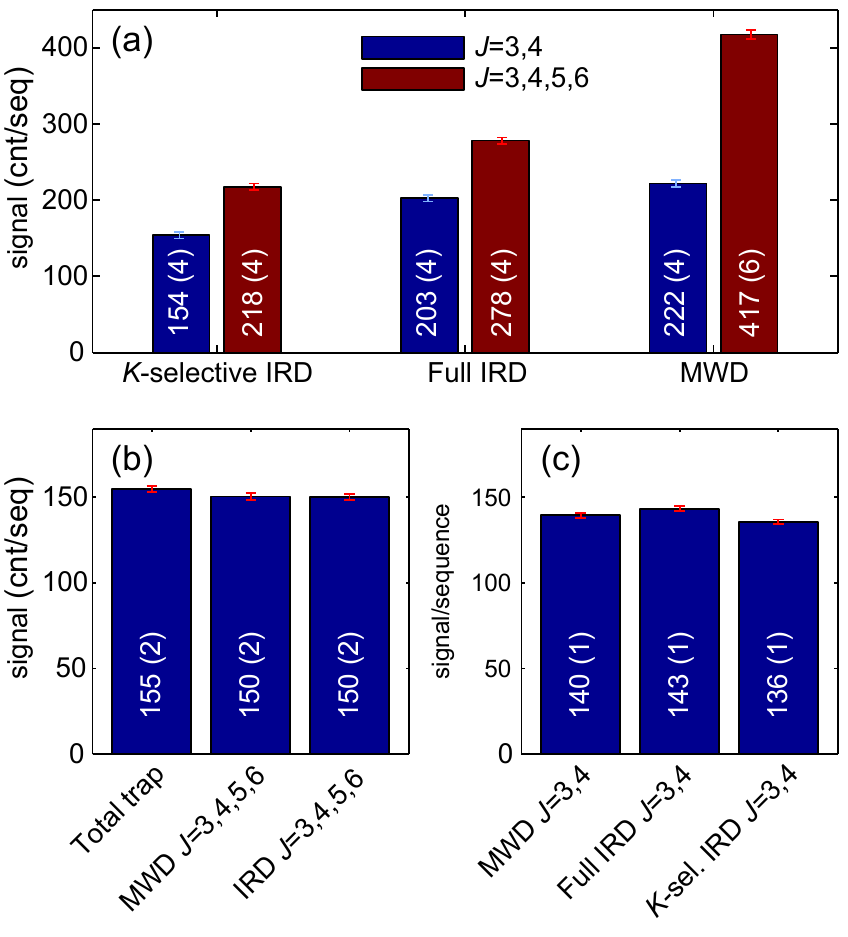}
\caption{Comparison of our detection methods. (a)~MWD, IRD and $K$-selective IRD applied to a warm ensemble in our trap where other $J$ and $K$ states are present. Note that the total signal of trapped molecules is about \unit[940]{cnt/seq}, if no depletion is applied. (b)~Proof that MWD and IRD deplete all molecules populating the states of interest. Here we use a molecular ensemble where no other $J$ and $K$ states are present (see main text). The detection is applied for \unit[8]{s} to ensure full saturation. (c)~Comparison of the $K$-selective IRD measurement with MWD and full IRD for an ensemble where no other $K$ states are present. The $K$-selective measurement underestimates the signal of molecules by less than 5\,\%.}
\label{fig:comparison}
\end{figure}

In this section we compare the state detection methods MWD and IRD focusing on the ability to discriminate rotational states and the ability to detect all molecules populating the states of interest. First, we compare the number of molecules detected in the $|3,4;  3;  M\rangle$ and the $|3,4,5,6;  3;  M\rangle$ states using either MWD or IRD. We use an ensemble directly loaded from the electric quadrupole guide closely matching the rotational state distribution shown in figure~\ref{fig:dist}. The molecules are stored for \unit[3]{s} and either the depletion or nothing is applied during the last \unit[2]{s}.

In figure~\ref{fig:comparison}(a) the resulting state selected signal of molecules is plotted for each method, satisfying our expectations from the previous discussions. In particular we see that the $|K|$-selective IRD measurement always results in the lowest signal as this methods underestimates the number of molecules in the states of interest. In contrast, MWD and full IRD overestimate the signal where the effect is stronger for MWD than for IRD due to the enhanced depletion of states with $|K|\ne3$. All three methods differ less for the depletion of the states with $J=3,4$ than for the states with $J=3,4,5,6$ as the contribution of $|K|\ne3$ is stronger for $J=5,6$ than for $J=3,4$.

To verify that we can deplete all molecules within a certain set of states we use an almost pure state ensemble produced by Sisyphus cooling. Due to the long trapping time during cooling, a reduced trap voltage for unloading, and the RF knife for cooling, almost only the cooled molecules (\unit[30]{mK}) in the $|3,4;  3;  M\rangle$ states are unloaded for detection at the end of the cooling sequence~\cite{Zeppenfeld2012a}. However, some molecules are pumped to the $J=5,6$ states due to blackbody radiation. In addition, a Fermi resonance with a double excited vibrational state can also lead to population of the $J=5$ state. We thus expect most of the molecules to populate the $|3,4;  3;  M\rangle$ states and some the $|5,6;  3;  M\rangle$ states. To test our state sensitive detection methods we used this cold ensemble. The depletion was applied to the $|3,4,5,6;3; M\rangle$ states for \unit[8]{s} to ensure full saturation. Figure~\ref{fig:comparison}(b) shows the outcome of this measurement. Here, both methods, MWD and full IRD, give the same result which is almost equivalent to the total signal of trapped molecules. We can therefore state that the depletion transfers all molecules populating the states of interest to untrapped states. A shorter depletion time leads to some error, however, already for two seconds of depletion this can most often be ignored.

We additionally used the Sisyphus-cooled ensemble to compare the $|K|$-selective IRD measurement with MWD and full IRD, as states with $|K|\ne3$ are hardly populated. We detected the population of the $|3,4;  3;  M\rangle$ states with \unit[2]{s} for depletion. Figure~\ref{fig:comparison}(c) shows that the $K$-selective IRD measurement underestimates the signal by less than 5\,\% compared to MWD and full IRD.

\subsection{Detection of the population in the $|4;  3;  4\rangle$ state}
\label{sec:Results.M}

\begin{figure}[t]
\centering
\includegraphics{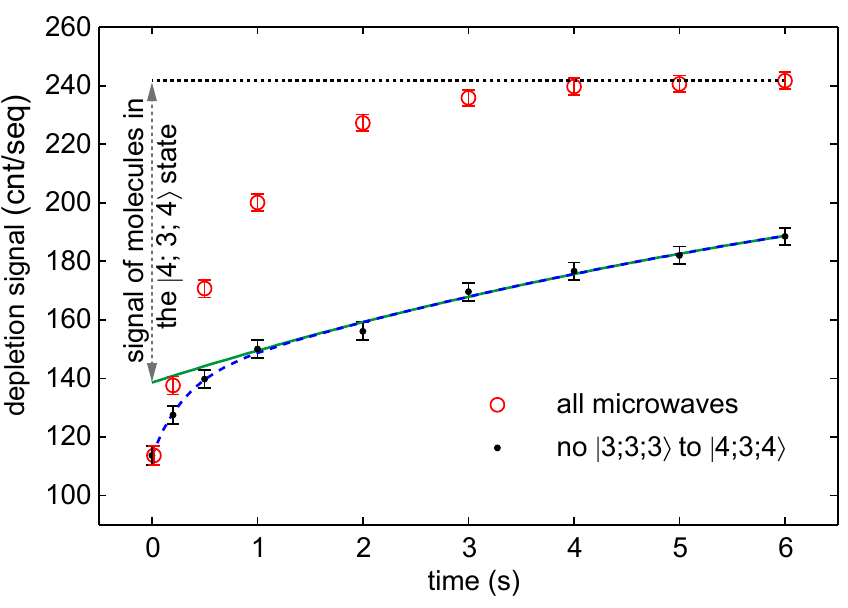}
\caption{
Detection of the population in the $|4;  3;  4\rangle$ state. The figure shows two saturation measurements. The red circles corresponds to an IRD measurement. The black dots are measured without driving the transition $|3;  3;  3\rangle \leftrightarrow |4;  3;  4\rangle$. The difference of the two measurements yields a measure for the number of molecules in the $|4;  3;  4\rangle$ state. Due to the Stark broadening of our spectral lines, the $|3;  3;  3\rangle \leftrightarrow |4;  3;  4\rangle$ transition is slightly driven by the other applied microwave frequencies. Measuring only at a single point with \unit[2]{s} of depletion therefore underestimates the number of molecules in the $|4;  3;  4\rangle$ state as is explained in the main text.
}     
\label{fig:Mdet}
\end{figure}

In this section, we discuss the experimental results of a single $M$-substate detection using the scheme presented in section~\ref{sec:Scheme.M}. As we have seen in section~\ref{sec:Rates}, the scheme with IRD is expected to work and was in fact already used to measure our electric field distribution. Here we again perform saturation measurements to obtain an estimation of the error of our single $M$-substate detection.

To improve statistics we used an ensemble cooled to \unit[150]{mK} (cf. section~\ref{sec:Limit}. Afterwards, the population was distributed among the $|4;  3;  M\rangle$ states. For the following \unit[6]{s} the molecules were stored, and we performed an IRD saturation measurement either with driving the $|3;  3;  3\rangle \leftrightarrow |4;  3;  4\rangle$ MW transition or without. To improve the quality of our single state detection we reduced the MW power by about a factor of eight compared to the IRD measurements described above. Note that the laser is left on during the whole \unit[6]{s}.

Figure~\ref{fig:Mdet} shows a clear difference in the signals. The measurement with applying the $|3;  3;  3\rangle \leftrightarrow |4;  3;  4\rangle$  coupling (red circles) is equivalent to the full IRD signal (except for the reduced MW power). The measurement without (black dots) has to be separated into two regions: the fast increase yields the depletion signal of all molecules populating the $|3,4;  3;  M\rangle$ states except the population of the  $|4;  3;  4\rangle$ state. The subsequent slower increase is caused by the unintended driving of the $|3;  3;  3\rangle \leftrightarrow |4;  3;  4\rangle$  transition in the inhomogeneous electric field regions as well as by blackbody induced population transfer. 

The difference of both measurements yields a signal proportional to the number of molecules in the state $|4;  3;  4\rangle$ and thus the population of a state described by a single $J,K$ and $M$ quantum number. Due to the unintended depletion of the $|4;  3;  4\rangle$ state in the black curve, this difference underestimates the population in the state $|4;  3;  4\rangle$. To obtain an estimation of the error we fitted the black data points with a double exponential (blue, dashed). The slow part of this fit is plotted (solid green) and the difference of the fully saturated red curve and the extrapolation of the green curve to \unit[0]{s} yields the signal of molecules in the $|4;  3;  4\rangle$ state. The fast increase saturates after about \unit[2]{s}, meaning that the entire population of all states but the $|4;  3;  4\rangle$ state is depleted. For \unit[2]{s} of single state detection we thus underestimate the number of molecules by approximately 20\,\%. This measurement shows that we can clearly distinguish between the $|4;  3;  4\rangle$ state and the other states in the $|3,4;  3;  M\rangle$ manifold. 
 
The $|3;  3;  3\rangle$ state can be detected with a similar scheme. The MW coupling stays the same but the vibrational driving is on the $J=4$ states which has two consequences: First, the vibrational decay transfers population to the $J=5$ state. Thus MWD or IRD for depleting the $J=5$ state is additionally needed. Second, the spontaneous decay from the vibrational excited state $|4;  3;  3\rangle$ transfers population to the target state $|3;  3;  3\rangle$ with a branching ratio of 5\,\% which gives an additional error of the measurement.

\section{Conclusion}
\label{sec:Out}
In summary, we have presented a detailed investigation of rotational state detection of trapped molecules based on depletion, driving rotational and vibrational transitions to transfer molecules to untrapped states. As suitable transitions can be found in any molecule, our technique is extremely general and should be applicable to all trappable molecule species. Moreover, with use of only a single microwave synthesizer and (optionally) a single infrared laser, our technique is simple to implement. An extension to other types of internal states is feasible. Detecting hyperfine states would require a sufficient spectral resolution, and detecting vibrational states could make use of different rotational constants in different states. We have thus achieved a versatile tool to investigate polyatomic molecules.

\section{References}

\providecommand{\newblock}{}

\end{document}